\documentstyle[nato]{crckapb}

\newcommand{\stt}{\small\tt}

\begin{opening}

\author{J.D. VERGADOS}

\title{SEARCHING FOR SUSY DARK MATTER \protect\\ 
      - THE DIRECTIONAL RATE AND THE MODUALTION EFFECT} 

\institute{
Theoretical Physics Division, University of Ioannina\\ 
Ioannina, GR-45110, Greece.}

\end{opening}

\runningtitle{SEARCHING FOR SUSY DARK MATTER} 

\begin{document}


\begin{abstract}
 The direct detection of  dark matter
is central to particle physics and cosmology. Current fashionable supersymmetric
models provide a natural dark matter candidate which is the lightest
supersymmetric particle (LSP). Such models combined with fairly well 
understood physics like
the quark substructure of the nucleon and the structure of the nucleus
(form factor 
and/or spin response function), permit the evaluation of
the event rate for LSP-nucleus elastic scattering. The thus obtained event rates
are, however, very low or even undetectable.
 So it is imperative to exploit characteristic signatures, like the modulation effect, i.e. the dependence of
the event rate on  the earth's annual motion, and 
the directional rate, i.e its dependence on the direction of the recoiling
nucleus. In this paper we do this 
using various velocity distributions, isothermal (symmetric as well
as only axially asymmetric) and non isothermal (e.g. due to caustic 
rings). 
\end{abstract}

\section{Introduction}
 The consideration of  exotic dark matter has become necessary
in order to close the Universe \cite {KTP}$^,$ \cite{Jungm}. Furthermore in
in order to understand the large scale structure of the universe 
one has to consider
 matter made up of particles which were 
non-relativistic at the time of freeze out, i.e  cold dark 
matter (CDM). The COBE data ~\cite{COBE} suggest that CDM
is at least $60\%$ ~\cite {GAW}. 
 Recent data from the 
High-z Supernova Search Team \cite{HSST} and
Supernova Cosmology Project  ~\cite {SPF} 
$^,$~\cite {SCP} suggest the necessity of the cosmological constant $\Lambda$. In fact
the situation can adequately be
described by  a barionic component $\Omega_B=0.1$ along with the exotic 
components $\Omega _{CDM}= 0.3$ and $\Omega _{\Lambda}= 0.6$.
In another analysis Turner \cite {Turner} gives 
$\Omega_{m}= \Omega _{CDM}+ \Omega _B=0.4$.
Since the non exotic component cannot exceed $40\%$ of the CDM 
~\cite{Jungm}$^,$~\cite {Benne}, there is room for the exotic WIMP's 
(Weakly  Interacting Massive Particles).
  In fact the DAMA experiment ~\cite {BERNA2} 
has claimed the observation of one signal in direct detection of a WIMP, which
with better statistics has subsequently been interpreted as a modulation signal
~\cite{BERNA1}.

The above developments are in line with particle physics considerations. Thus,
in the currently favored supersymmetric (SUSY)
extensions of the standard model, the most natural WIMP candidate is the LSP,
i.e. the lightest supersymmetric particle. In the most favored scenarios the
LSP can be simply described as a Majorana fermion, a linear 
combination of the neutral components of the gauginos and Higgsinos
\cite{Jungm}$^,$~\cite{JDV}$^-$\cite{Hab-Ka}. 

 Since this particle is expected to be very massive, $m_{\chi} \geq 30 GeV$, and
extremely non relativistic with average kinetic energy $T \leq 100 KeV$,
it can be directly detected ~\cite{JDV}$^-$\cite{KVprd} mainly via the recoiling
of a nucleus (A,Z) in the elastic scattering process:
\begin{equation}
\chi \, +\, (A,Z) \, \to \, \chi \,  + \, (A,Z)^* 
\end{equation}
($\chi$ denotes the LSP). In order to compute the event rate one needs
the following ingredients:

1) An effective Lagrangian at the elementary particle 
(quark) level obtained in the framework of supersymmetry as described 
in Refs.~\cite{Jungm}, Bottino {\it et al.} \cite{ref2}, 
\cite{Hab-Ka},
Kane {\it et al.}, Castano {\it et al.} and Arnowitt {\it et al}.
 Our own SUSY input parameters will appear elsewhere \cite {Gomez}
:
2) A procedure in going from the quark to the nucleon level, i.e. a quark 
model for the nucleon. The results, for the scalar and the isoscalar
axial couplings, depend crucially on the content of the
nucleon in quarks other than u and d ~\cite{Dree,Chen}.

3) Compute the relevant nuclear matrix elements~
\cite{Ress}$^-$\cite{DIVA00}.

Since the obtained rates are very low, one would like to be able to exploit the
modulation of the event rates due to the earth's
revolution around the sun. 
One also would like to know the directional rates, by observing the 
nucleus in a certain direction, which correlate with the motion of the
sun around the center of the galaxy.

 The purpose of our present review is to focus on this last point 
along the lines suggested by our recent work \cite{JDV99,JDV99b}.

\section{Expressions for the Unconvoluted Event Rates.} 
Combining for results of the previous section we can write
\begin{eqnarray}
{\it L}_{eff} = - \frac {G_F}{\sqrt 2} \{({\bar \chi}_1 \gamma^{\lambda}
\gamma_5 \chi_1) J_{\lambda} + ({\bar \chi}_1 
 \chi_1) J\}
 \label{eq:eg 41}
\end{eqnarray}
where
\begin{eqnarray}
  J_{\lambda} =  {\bar N} \gamma_{\lambda} (f^0_V +f^1_V \tau_3
+ f^0_A\gamma_5 + f^1_A\gamma_5 \tau_3)N~~,~~
J = {\bar N} (f^0_s +f^1_s \tau_3) N
 \label{eq:eg.42}
\end{eqnarray}

We have neglected the uninteresting pseudoscalar and tensor
currents. Note that, due to the Majorana nature of the LSP, 
${\bar \chi_1} \gamma^{\lambda} \chi_1 =0$ (identically).

 With the above ingredients the differential cross section can be cast in the 
form 
\begin{equation}
d\sigma (u,\upsilon)= \frac{du}{2 (\mu _r b\upsilon )^2} [(\bar{\Sigma} _{S} 
                   +\bar{\Sigma} _{V}~ \frac{\upsilon^2}{c^2})~F^2(u)
                       +\bar{\Sigma} _{spin} F_{11}(u)]
\label{2.9}
\end{equation}
\begin{equation}
\bar{\Sigma} _{S} = \sigma_0 (\frac{\mu_r}{m_N})^2  \,
 \{ A^2 \, [ (f^0_S - f^1_S \frac{A-2 Z}{A})^2 \, ] \simeq \sigma^S_{p,\chi^0}
        A^2 (\frac{\mu_r}{m_N})^2 
\label{2.10}
\end{equation}
\begin{equation}
\bar{\Sigma} _{spin}  =  \sigma^{spin}_{p,\chi^0}~\zeta_{spin}
\label{2.10a}
\end{equation}
\begin{equation}
\zeta_{spin}= \frac{(\mu_r/m_N)^2}{3(1+\frac{f^0_A}{f^1_A})^2}
[(\frac{f^0_A}{f^1_A} \Omega_0(0))^2 \frac{F_{00}(u)}{F_{11}(u)}
  +  2\frac{f^0_A}{ f^1_A} \Omega_0(0) \Omega_1(0)
\frac{F_{01}(u)}{F_{11}(u)}+  \Omega_1(0))^2  \, ] 
\label{2.10b}
\end{equation}
\begin{equation}
\bar{\Sigma} _{V}  =  \sigma^V_{p,\chi^0}~\zeta_V 
\label{2.10c}
\end{equation}
\begin{equation}
\zeta_V = \frac{(\mu_r/m_N)^2} {(1+\frac{f^1_V}{f^0_V})^2} A^2 \, 
(1-\frac{f^1_V}{f^0_V}~\frac{A-2 Z}{A})^2 [ (\frac{\upsilon_0} {c})^2  
[ 1  -\frac{1}{(2 \mu _r b)^2} \frac{2\eta +1}{(1+\eta)^2} 
\frac{\langle~2u~ \rangle}{\langle~\upsilon ^2~\rangle}] 
\label{2.10d}
\end{equation}
\\
$\sigma^i_{p,\chi^0}=$ proton cross-section,$i=S,spin,V$ given by:\\
$\sigma^S_{p,\chi^0}= \sigma_0 ~(f^0_S)^2$   (scalar) , 
(the isovector scalar is negligible, i.e. $\sigma_p^S=\sigma_n^S)$\\
$\sigma^{spin}_{p,\chi^0}= \sigma_0~~3~(f^0_A+f^1_A)^2$
  (spin) ,
$\sigma^{V}_{p,\chi^0}= \sigma_0~(f^0_V+f^1_V)^2$  
(vector)   \\
where $m_p$ is the proton mass,
 $\eta = m_x/m_N A$, and
 $\mu_r$ is the reduced mass and  
\begin{equation}
\sigma_0 = \frac{1}{2\pi} (G_F m_N)^2 \simeq 0.77 \times 10^{-38}cm^2 
\label{2.7} 
\end{equation}
\begin{equation}
u = q^2b^2/2~~or~~
Q=Q_{0}u, \qquad Q_{0} = \frac{1}{A m_{N} b^2} 
\label{2.15} 
\end{equation}
where
b is (the harmonic oscillator) size parameter, 
q is the momentum transfer to the nucleus, and
Q is the energy transfer to the nucleus\\
In the above expressions $F(u)$ is the nuclear form factor and
$F_{\rho \rho^{\prime}}(u)$
are the spin form factors \cite{KVprd} ($\rho , \rho^{'}$ are isospin indices)\\
Both form factors are normalized to one at $u=0$.\\

 The nom-directional event rate is given by:
\begin{equation}
R=R_{non-dir} =\frac{dN}{dt} =\frac{\rho (0)}{m_{\chi}} \frac{m}{A m_N} 
\sigma (u,\upsilon) | {\boldmath \upsilon}|
\label{2.17} 
\end{equation}
 Where
 $\rho (0) = 0.3 GeV/cm^3$ is the LSP density in our vicinity and 
 m is the detector mass 
The differential non-directional  rate can be written as
\begin{equation}
dR= \frac{\rho (0)}{m_{\chi}} \frac{m}{A m_N} 
d\sigma (u,\upsilon) | {\boldmath \upsilon}|
\label{2.18}  
\end{equation}
where $d\sigma(u,\upsilon )$ was given above.

 The directional differential rate \cite{Copi99} in the direction
$\hat{e}$ is 
given by :
\begin{equation}
dR_{dir} = \frac{\rho (0)}{m_{\chi}} \frac{m}{A m_N} 
{\boldmath \upsilon}.\hat{e} H({\boldmath \upsilon}.\hat{e})
 ~\frac{1}{2 \pi}~  
d\sigma (u,\upsilon)
\label{2.20}  
\end{equation}
where H is the Heaviside step function. The factor of $1/2 \pi$ is 
introduced, since we have chosen to normalize our results to the
usual differential rate.

\section{Convolution of the Event Rate}
 We have seen that the event rate for LSP-nucleus scattering depends on the
relative LSP-target velocity. In this section we will examine the consequences 
of the earth's
revolution around the sun (the effect of its rotation around its axis is
expected to
be negligible) i.e. the modulation effect. In practice this has been 
accomplished by
assuming a consistent LSP velocity dispersion, such as  
a Maxwell distribution ~\cite{Jungm} or its
extensions with only axial symmetry ~\cite{Druk}, already
been discussed in the lierature \cite{JDV99,JDV99b}.
More recently other non-isothermal
approaches, in the context velocity peaks and caustic rings, have been 
proposed, see e.g Sikivie
et al \cite{SIKIVIE}. 

 In the last case one considers the late in-fall of dark matter
into our  galaxy producing flows of caustic rings. In particular the predictions of a self-similar model have been put forward as a possible scenario
for dark matter density-velocity distribution, see e.g. Sikivie et al
\cite{SIKIVIE}. The implications of such theoretical schemes and,
in particular, the modulation effect are the subject of this section. 

 Following Sikivie we will consider $2 \times N$ caustic rings, (i,n)
, i=(+.-) and n=1,2,...N (N=20 in the model of Sikivie et al),
each of which
contributes to the local density a fraction $\bar{\rho}_n$ of the
of the summed density $\bar{\rho}$ of each of the $i=+,-$. It contains
WIMP like particles with velocity 
${\bf y}^{'}_n=(y^{'}_{nx},y^{'}_{ny},y^{'}_{nz})$
in units of essentially
the sun's velocity ($\upsilon_0=220~Km/s$), with respect to the
galactic center.
The z-axis is chosen in the direction of the disc's rotation,
 i.e. in the direction of the motion of the
the sun, the y-axis is perpendicular to the plane of the galaxy and 
the x-axis is in the radial direction. We caution the reader that
these axes are traditionally indicated by astronomers as
$\hat{e}_{\phi},\hat{e}_r, \hat{e}_z)$ respectively. 
The needed quantities are taken from the 
work of Sikivie et al \cite{SIKIVIE}, via the 
definitions

$y^{'}_n=\upsilon_n/\upsilon_0
,y^{'}_{nz}=\upsilon_{n\phi}/\upsilon_0=y_{nz}
,y^{'}_{nx}=\upsilon_{nr}/\upsilon_0=y_{nx}
,y^{'}_{ny}=\upsilon_{nz}/\upsilon_0=y_{ny}
,\rho_{n}=d_n/\bar{\rho}
,\bar{\rho}=\sum_{n=1}^N~d_n$ and 
$y_n=[(y_{nz}-1)^2+y^2_{ny}+y^2_{nx}]^{1/2}$ (for each flow +.-).
 This leads to a
velocity distribution of the form:
\begin{eqnarray}
f(\upsilon^{\prime}) = \sum_{n=1}^N~\delta({\bf \upsilon} ^{'}
    -\upsilon_0~{\bf y}^{'}_n)
\label{3.1}  
\end{eqnarray}
Since the axis of the ecliptic \cite{KVprd}. 
lies very close to the $y,z$ plane the velocity of the earth around the
sun is given by 
\begin{eqnarray}
{\bf \upsilon}_E \, = \, {\bf \upsilon}_0 \, + \, {\bf \upsilon}_1 \, 
= \, {\bf \upsilon}_0 + \upsilon_1(\, sin{\alpha} \, {\bf \hat x}
-cos {\alpha} \, cos{\gamma} \, {\bf \hat y}
+cos {\alpha} \, sin{\gamma} \, {\bf \hat z} \,)
\label{3.6}  
\end{eqnarray}
where $\alpha$ is the phase of the earth's orbital motion, 
$\alpha =0$ around second of June.

One can now express the above distribution in the laboratory frame 
\cite{JDV99b}
by writing $ {\bf \upsilon}^{'}={\bf \upsilon} \, + \, {\bf \upsilon}_E \,$ 

\section{Expressions for the Differential   Event Rate}
The mean value of the non-directional event rate of Eq. (\ref {2.18}), 
is given by
\begin{eqnarray}
\Big<\frac{dR}{du}\Big> =\frac{\rho (0)}{m_{\chi}} 
\frac{m}{A m_N}  
\int f({\bf \upsilon}, {\boldmath \upsilon}_E) 
          | {\boldmath \upsilon}|
                       \frac{d\sigma (u,\upsilon )}{du} d^3 {\boldmath \upsilon} 
\label{3.10} 
\end{eqnarray}
 The above expression can be more conveniently written as
\begin{eqnarray}
\Big<\frac{dR}{du}\Big> =\frac{\rho (0)}{m_{\chi}} \frac{m}{Am_N} \sqrt{\langle
\upsilon^2\rangle } {\langle \frac{d\Sigma}{du}\rangle } 
\label{3.11}  
\end{eqnarray}
where
\begin{eqnarray}
\langle \frac{d\Sigma}{du}\rangle =\int
           \frac{   |{\boldmath \upsilon}|}
{\sqrt{ \langle \upsilon^2 \rangle}} f({\boldmath \upsilon}, 
         {\boldmath \upsilon}_E)
                       \frac{d\sigma (u,\upsilon )}{du} d^3 {\boldmath \upsilon}
\label{3.12}  
\end{eqnarray}

 There are now experiments under way aiming at measuring directional rates
, i.e. the case in which the nucleus is observed in a certain direction.
The rate will depend on the direction of observation, showing a strong
correlation with the direction of the sun's motion. In a favorable 
situation the rate will merely be suppressed by a factor of $2 \pi$
relative to the non-directional rate. This is due to the fact that one 
does not now integrate over the
azimuthal angle of the nuclear recoiling momentum. The directional rate
will also show modulation due to the Earth's motion. 

The mean value of the directional differential event rate of Eq. (\ref {2.20}), 
is defined by
\begin{eqnarray}
\Big<\frac{dR}{du}\Big>_{dir} =\frac{\rho (0)}{m_{\chi}} 
\frac{m}{A m_N} \frac{1}{2 \pi} 
\int f({\bf \upsilon}, {\boldmath \upsilon}_E)
{\boldmath \upsilon}.\hat{e} H({\boldmath \upsilon}.\hat{e})
                       \frac{d\sigma (u,\upsilon )}{du} d^3 {\boldmath \upsilon} 
\label{4.10} 
\end{eqnarray}
It can be more conveniently expressed as
\begin{eqnarray}
\Big<\frac{dR}{du}\Big>_{dir} =\frac{\rho (0)}{m_{\chi}} \frac{m}{Am_N} \sqrt{\langle
\upsilon^2\rangle } {\langle \frac{d\Sigma}{du}\rangle }_{dir} 
\label{4.11}  
\end{eqnarray}
where
\begin{eqnarray}
\langle \frac{d\Sigma}{du}\rangle _{dir}=\frac{1}{2 \pi} \int \frac{ 
{\boldmath \upsilon}.\hat{e} H({\boldmath \upsilon}.\hat{e})}
{\sqrt{ \langle \upsilon^2 \rangle}} f({\boldmath \upsilon}, {\boldmath \upsilon}_E)
                       \frac{d\sigma (u,\upsilon )}{du} d^3 {\boldmath \upsilon}
\label{4.12}  
\end{eqnarray} 
We are not going to discuss the differential rates further.
We will limit our discussion to the case of the total rates.

\section{The Total  non-directional Event Rates}
Integrating the differential rate  from $u_{min}$ to $u_{max}$ we
 obtain for the total non-directional ratei. In the case of caustic rings
we find:
\begin{eqnarray}
R =  \bar{R}\, t \, \frac{2 \bar{\rho}}{\rho(0)}
          [1 - h(a,Q_{min})cos{\alpha})] 
\label{3.55a}  
\end{eqnarray}
where
\begin{eqnarray}
\bar{R} =\frac{\rho (0)}{m_{\chi}} \frac{m}{Am_N} \sqrt{\langle
v^2\rangle } [\bar{\Sigma}_{S}+ \bar{\Sigma} _{spin} + 
\frac{\langle \upsilon ^2 \rangle}{c^2} \bar{\Sigma} _{V}]
\label{3.39b}  
\end{eqnarray}
with $\bar{\Sigma} _{i}, i=S,V,spin$ have been defined above, see Eqs
 (\ref {2.10}) - (\ref {2.10c}). Furthermore 
$a = [\sqrt{2} \mu _rb\upsilon _0]^{-1}$  
\begin{eqnarray}
u_{min}= \frac{Q_{min}}{Q_0},
u_{max}=min(\frac{y^2_{esc}}{a^2},max(\frac{y_{n} ^2}{a^2})~,~ n=1,2,...,N)
\label{3.30c}  
\end{eqnarray}
Here $y_{esc}=\frac{\upsilon_{esc}}{\upsilon_0}$, with 
$\upsilon_{escape}=625 Km/s$
is the escape velocity from the galaxy.
 The modulation is described by the parameter $h$. 
 Similarly integrating Eq. (\ref {3.12}) we obtain for the total
non-directional rate in our isothermal model as follows:
\begin{eqnarray}
R =  \bar{R}\, t \, [(1 + h(a,Q_{min})cos{\alpha})] 
\label{3.55b}  
\end{eqnarray}
Now $u_{max}=y_{esc} ^2/{a^2}$ 
Note the difference of sign in the definition of the modulation amplitude h 
compared to Eq. (\ref {3.55a}). 
 The modulation can be described in terms of the parameter $h$. 
 The effect of folding
with LSP velocity on the total rate is taken into account via the quantity
$t$. 

The meaning of $t$ is clear from the above discussion. We only like to
stress that it is a common practice to extract the LSP nucleon cross 
section from the  the expected experimental event rates in order to 
compare with the SUSY predictions as a function of the LSP mass. 

\section{The Total  Directional Event Rates}
 We will again examine the case of caustic rings and the isothermal
models considered above.

\subsection{The Total  Directional Event Rates in Non-isothermal Models}
 We need distinguish  
distinguish the following cases: a) $\hat{e}$ has a
component in the sun's direction of
motion, i.e. $0<\theta < \pi /2$, labeled by i=u (up). b) Detection
in the direction specified by  $\pi /2 <\theta < \pi $, labeled by 
i=d (down). Thus :

1. In the first quadrant (azimuthal angle $0 \leq \phi \leq \pi/2)$.
\begin{eqnarray}
R^i_{dir} & = &\bar{R} \frac{2 \bar{\rho}}{\rho(0)}
    \frac{t}{2 \pi} [(r^i_z  - \cos \alpha~ h^i_1) {\bf e}_z.{\bf e}  
\nonumber \\ 
&+& (r^i_y +cos \alpha h^i_2 +\frac{h^i_c }{2}(|cos\alpha|+cos\alpha))
     |{\bf e}_y.{\bf e} | 
\nonumber \\ 
&+& (r^i_x -sin \alpha h^i_3 +\frac{h^i_s }{2}(|sin\alpha|-sin\alpha))
     |{\bf e}_x.{\bf e} | ]
\label{3.56}  
\end{eqnarray}
2. In the second quadrant (azimuthal angle $\pi/2 \leq \phi \leq \pi)$
\begin{eqnarray}
R^i_{dir} & = &\bar{R} \frac{2 \bar{\rho}}{\rho(0)}
    \frac{t}{2 \pi}  [(r^i_z  - \cos \alpha~ h^i_1) {\bf e}_z.{\bf e}  
\nonumber \\ 
&+& (r^i_y +cos \alpha h^i_2 (u)+\frac{h^i_c }{2}(|cos\alpha|-cos\alpha))
     |{\bf e}_y.{\bf e} | 
\nonumber \\ 
&+& (r^i_x +sin \alpha h^i_3 +\frac{h^i_s }{2}(|sin\alpha|+sin\alpha))
     |{\bf e}_x.{\bf e} | ]
\label{3.57}  
\end{eqnarray}
3. In the third quadrant (azimuthal angle $\pi \leq \phi \leq 3 \pi/2)$.
\begin{eqnarray}
R^i_{dir} & = &\bar{R} \frac{2 \bar{\rho}}{\rho(0)}
    \frac{t}{2 \pi}  [(r^i_z  - \cos \alpha~ h^i_1) {\bf e}_z.{\bf e}  
\nonumber \\ 
&+& (r^i_y -cos \alpha h^i_2 (u)+\frac{h^i_c (u)}{2}(|cos\alpha|-cos\alpha))
     |{\bf e}_y.{\bf e} | 
\nonumber \\ 
&+& (r^i_x +sin \alpha H^i_3 +\frac{h^i_s }{2}(|sin\alpha|+sin\alpha))
     |{\bf e}_x.{\bf e} | ]
\label{3.58}  
\end{eqnarray}
4. In the fourth quadrant (azimuthal angle $3 \pi/2 \leq \phi \leq 2 \pi)$
\begin{eqnarray}
R^i_{dir} & = &\bar{R} \frac{2 \bar{\rho}}{\rho(0)}
    \frac{t}{2 \pi}  [(r^i_z  - \cos \alpha~ h^i_1) {\bf e}_z.{\bf e}  
\nonumber \\ 
&+& (r^i_y -cos \alpha h^i_2 +\frac{h^i_c }{2}(|cos\alpha|-cos\alpha))
     |{\bf e}_y.{\bf e} | 
\nonumber \\ 
&+& (r^i_x -sin \alpha h^i_3 +\frac{h^i_s }{2}(|sin\alpha|-sin\alpha))
     |{\bf e}_x.{\bf e} | ]
\label{3.59 }  
\end{eqnarray}

\subsection{The Total  Directional Event Rates in Isothermal Models}
 In this case we find it convenient to consider 
the absolute value of the difference of the
rates in the two opposite directions.
Integrating Eq. (\ref {4.10}) and restricting ourselves close to the 
axes we obtain
\begin{eqnarray}
R_{dir}& = &  \bar{R}\, (t^0/4 \pi) \, 
                  |(1 + h_1(a,Q_{min})cos{\alpha}){\bf e}~_z.{\bf e}
\nonumber\\  &-& h_2(a,Q_{min})\, 
cos{\alpha} {\bf e}~_y.{\bf e}
                      + h_3(a,Q_{min})\, 
sin{\alpha} {\bf e}~_x.{\bf e}|
\label{4.55}  
\end{eqnarray}
Note that now the rate is  normalized to $t^0/2$ and not to $t$ and
that the modulation can
be described in terms of three parameters $h_l$, l=1,2,3. 

\section{Results and Discussion}

 We will discuss the effects of folding with the LSP velocity combined with the
nuclear form factor and specialized in the case of the nucleus
$^{127}I$, 
which is one of the most popular targets \cite{BERNA2}$^,$\cite{Smith}
\cite{Primack}$^-$ \cite {Verg98}for the enegy cutoffs:
$Q_{min}=0,~10,~20$ KeV.  
 
 In the case of the non isothermal model of Sikivie et al 
\cite{SIKIVIE}, 
the total rates are 
described in terms of the quantities $t,r^i_x,r^i_y,r^i_z$ for the
unmodulated amplitude  and $h,h^i_1,h^i_2,h^i_3,h^i_c,h^i_s$ $i=u,d$
for the modulated one. 
\begin{table}[htb]
\begin{center}
\caption{The quantities $t$ and $h$ entering the total non-directional
rate in the case of the
target $_{53}I^{127}$ for various LSP masses and $Q_{min}$ in KeV. 
Also shown are the quantities $r^i_j,h^i_j$
 $i=u,d$ and $j=x,y,z,c,s$, entering the directional rate for no energy
cutoff. For definitions see text. 
}
\begin{tabular}{|c|c|rrrrrrr|}
\hline
& & & & & & & &     \\
&  & \multicolumn{7}{|c|}{LSP \hspace {.2cm} mass \hspace {.2cm} in GeV}  \\ 
\hline 
& & & & & & & &     \\
Quantity &  $Q_{min}$  & 10  & 30  & 50  & 80 & 100 & 125 & 250   \\
\hline 
& & & & & & & &     \\
t      &0.0&1.451& 1.072& 0.751& 0.477& 0.379& 0.303& 0.173\\
h      &0.0&0.022& 0.023& 0.024& 0.025& 0.026& 0.026& 0.026\\
$r^u_z$&0.0&0.726& 0.737& 0.747& 0.757& 0.760& 0.761& 0.761\\
$r^u_y$&0.0&0.246& 0.231& 0.219& 0.211& 0.209& 0.208& 0.208\\
$r^u_x$&0.0&0.335& 0.351& 0.366& 0.377& 0.380& 0.381& 0.381\\
$h^u_z$&0.0&0.026& 0.027& 0.028& 0.029& 0.029& 0.030& 0.030\\
$h^u_y$&0.0&0.021& 0.021& 0.020& 0.020& 0.019& 0.019& 0.019\\
$h^u_x$&0.0&0.041& 0.044& 0.046& 0.048& 0.048& 0.049& 0.049\\
$h^u_c$&0.0&0.036& 0.038& 0.040& 0.041& 0.042& 0.042& 0.042\\
$h^u_s$&0.0&0.036& 0.024& 0.024& 0.023& 0.023& 0.022& 0.022\\
$r^d_z$&0.0&0.274& 0.263& 0.253& 0.243& 0.240& 0.239& 0.239\\
$r^d_y$&0.0&0.019& 0.011& 0.008& 0.007& 0.007& 0.007& 0.007\\
$r^d_x$&0.0&0.245& 0.243& 0.236& 0.227& 0.225& 0.223& 0.223\\
$h^d_z$&0.0&0.004& 0.004& 0.004& 0.004& 0.004& 0.004& 0.004\\
$h^d_y$&0.0&0.001& 0.000& 0.000& 0.000& 0.000& 0.000& 0.000\\
$h^d_x$&0.0&0.022& 0.021& 0.021& 0.020& 0.020& 0.020& 0.020\\
$h^d_c$&0.0&0.019& 0.018& 0.018& 0.017& 0.017& 0.017& 0.017\\
$h^d_s$&0.0&0.001& 0.001& 0.000& 0.000& 0.000& 0.000& 0.000\\
\hline 
& & & & & & & &     \\
t    &10.0&0.000& 0.226& 0.356& 0.265& 0.224& 0.172& 0.098\\
h    &10.0&0.000& 0.013& 0.023& 0.025& 0.025& 0.026& 0.026\\
\hline 
& & & & & & & &     \\
t    &20.0&0.000& 0.013& 0.126& 0.139& 0.116& 0.095& 0.054\\
h    &20.0&0.000& 0.005& 0.017& 0.024& 0.025& 0.026& 0.026\\
\hline
\end{tabular}
\end{center}
\end{table}
(see Table 1). Out of this list list only  
$t$ and $h$
enter the non-directional rate. We notice that the usual
modulation amplitude $h$ is smaller than the one arising in isothermal
models \cite {JDV99,JDV99b}. 
 As expected, the parameter t decreases as the reduced mass increases.
 
  In the case of isothermal models we will limit ourselves to the discussion
of the directional rates. If the direction of observation 
is  close to the coordinate axes, the rate is described in terms of the three
quantities $t_0$ and $h_i,~i=1,2,3$ (see Eq.  (\ref{4.55})). 
These are shown in tables 2-4 (for the differential rate see  
our previous work \cite {JDV99,JDV99b}.

\begin{table}[htb]
\begin{center}
\caption{The quantities $t^{0},h_1$ and $h_m$ for $\lambda=0$ in the case of the
target $_{53}I^{127}$ for various LSP masses and $Q_{min}$ in KeV (for
definitions see text). Only the scalar contribution is considered. Note that in
this case $h_2$ and $h_3$ are constants equal to 0.117 and 0.135 respectively.}
\begin{tabular}{|l|c|rrrrrrr|}
\hline
& & & & & & & &     \\
&  & \multicolumn{7}{|c|}{LSP \hspace {.2cm} mass \hspace {.2cm} in GeV}  \\ 
\hline 
& & & & & & & &     \\
Quantity &  $Q_{min}$  & 10  & 30  & 50  & 80 & 100 & 125 & 250   \\
\hline 
& & & & & & & &     \\
$t^0$ &0.0&1.960&1.355&0.886&0.552&0.442&0.360&0.212 \\
$h_1$ &0.0&0.059&0.048&0.037&0.029&0.027&0.025&0.023 \\
\hline 
& & & & & & & &     \\
$ t^0$ &10.&0.000&0.365&0.383&0.280&0.233&0.194&0.119 \\
$h_1$ &10.&0.000&0.086&0.054&0.038&0.033&0.030&0.025 \\
\hline 
& & & & & & & &     \\
$t^0$ &20.&0.000&0.080&0.153&0.136&0.11&0.102&0.065 \\
$h_1$ &20.&0.000&0.123&0.073&0.048&0.041&0.036&0.028 \\
\hline
\hline
\end{tabular}
\end{center}
\end{table}
\begin{table}[htb]
\begin{center}
\caption{The same as in the previous table, but for the value of the asymmetry 
parameter $\lambda=0.5$.}
\begin{tabular}{|l|c|rrrrrrr|}
\hline
& & & & & & & &     \\
&  & \multicolumn{7}{|c|}{LSP \hspace {.2cm} mass \hspace {.2cm} in GeV}  \\ 
\hline 
& & & & & & & &     \\
Quantity &  $Q_{min}$  & 10  & 30  & 50  & 80 & 100 & 125 & 250   \\
\hline 
& & & & & & & &     \\
$t^0$ & 0.0 &2.309&1.682&1.153&0.737&0.595&0.485&0.288 \\
$h_1$ & 0.0 &0.138&0.128&0.117&0.108&0.105&0.103&0.100\\
$h_2$ & 0.0 &0.139&0.137&0.135&0.133&0.133&0.133&0.132\\
$h_3$ & 0.0 &0.175&0.171&0.167&0.165&0.163&0.162&0.162\\
\hline 
& & & & & & & &     \\
$t^0$ & 10. &0.000&0.376&0.468&0.365&0.308&0.259&0.160\\
$h_1$ & 10. &0.000&0.174&0.139&0.120&0.114&0.110&0.103\\
$h_2$ & 10. &0.000&0.145&0.138&0.135&0.134&0.134&0.133\\
$h_3$ & 10. &0.000&0.188&0.174&0.167&0.165&0.164&0.162\\
\hline 
& & & & & & & &     \\
$t^0$ & 20. &0.000&0.063&0.170&0.171&0.153&0.134&0.087\\
$h_1$ & 20. &0.000&0.216&0.162&0.133&0.124&0.118&0.107\\
$h_2$ & 20. &0.000&0.155&0.143&0.137&0.136&0.135&0.133\\
$h_3$ & 20. &0.000&0.209&0.182&0.171&0.168&0.166&0.164\\
\hline
\end{tabular}
\end{center}
\end{table}
\begin{table}[htb]
\begin{center}
\caption{The same as in the previous, but for the value of the 
asymmetry parameter $\lambda=1.0$.}
\begin{tabular}{|l|c|rrrrrrr|}
\hline
& & & & & & & &     \\
&  & \multicolumn{7}{|c|}{LSP \hspace {.2cm} mass \hspace {.2cm} in GeV}  \\ 
\hline 
& & & & & & & &     \\
Quantity &  $Q_{min}$  & 10  & 30  & 50  & 80 & 100 & 125 & 250   \\
\hline 
& & & & & & & &     \\
$t^0$ & 0.0 &2.429&1.825&1.290&0.837&0.678&0.554&0.330\\
$h_1$ & 0.0 &0.192&0.182&0.170&0.159&0.156&0.154&0.150 \\
$h_2$ & 0.0 &0.146&0.144&0.141&0.139&0.139&0.138&0.138\\
$h_3$ & 0.0 &0.232&0.222&0.211&0.204&0.202&0.200&0.198\\
\hline 
& & & & & & & &     \\
$t^0$ & 10. &0.000&0.354&0.502&0.410&0.349&0.295&0.184\\
$h_1$ & 10. &0.000&0.241&0.197&0.174&0.167&0.162&0.154\\
$h_2$ & 10. &0.000&0.157&0.146&0.142&0.140&0.140&0.138\\
$h_3$ & 10. &0.000&0.273&0.231&0.213&0.208&0.205&0.200\\
\hline 
& & & & & & & &     \\
$t^0$ & 20. &0.000&0.047&0.169&0.186&0.170&0.150&0.100\\
$h_1$ & 20. &0.000&0.297&0.226&0.190&0.179&0.172&0.159\\
$h_2$ & 20. &0.000&0.177&0.153&0.144&0.142&0.141&0.139\\
$h_3$ & 20. &0.000&0.349&0.256&0.224&0.216&0.211&0.203\\
\hline
\end{tabular}
\end{center}
\end{table}

\section{Conclusions}
 We have studied the parameters describing 
the event rates for direct detection of SUSY dark matter and,
in particular, two 
characteristic signatures, which will aid the experimentalists in reducing
background: The non directional modulated event rates, which are
correlated with the motion of the Earth  and the directional event rates,
which are correlated with both the velocity of the sun
and the velocity of the Earth.

In the case of non isothermal models
\cite{SIKIVIE} from Table 1 we see that the maximun no longer 
occurs around June 2nd, but about six months later. The difference
between the maximum and the minimum is about $4\%$, smaller
than that predicted by the symmetric isothermal models 
\cite{JDV99,JDV99b}.  
In the case of the directional rate we found that the rates depend on the 
direction of observation. The biggest rates are obtained, If the
observation is made close to the direction of the sun's motion.
The directional rates are suppressed compared to the usual 
non-directional rates by the factor 
$f_{red}=\kappa/(2 \pi)$. We find that $\kappa=r^u_z \simeq 0.7$, 
if the observation is made in the sun's direction of motion, while
$\kappa\simeq 0.3$ in the opposite direction.
The modulation is a bit larger than in the non-directional case, but the
largest value, 
$8\%$, is not obtainrd along the sun's direction of motion, but 
in the x-direction (galactocentric direction).

 In the case of the isothermal models we restricted our discussion to the 
directional event rates. The reduction factor 
is now given by the parameter $f_{red}=t_0/(4 \pi~t)=\kappa/(2 \pi)$. We find
that $\kappa$ is around 0.6 for no asymmetry and around 0.7 for maximum
asymmetry ($\lambda=1.0$). In other words it is not very different from the 
naively expected $f_{red}=1/( 2 \pi)$. The 
modulation of the directional rate 
increases with the asymmetry parameter $\lambda$ and it depends, of
course, on
 the direction of observation. For $Q_{min}= 0$ it can reach values up 
to $23\%$. Larger values, up to $35\%$, are possible for large values of
 $Q_{min}$, but they occur at the expense of the total
number of counts.

 In all cases our results indicate that $t$
for large reduced mass deviates from unity.
Thus, if one  attempts to extract the LSP-nucleon cross section from
the data to compare it with the predictions of SUSY models,
one must take $t$ into account. 

{\it Acknowledgments:} One of the authors (JDV) would like to
acknowledge partial support of this work by $\Pi $ENE$\Delta $ 1895/95 of the 
Greek Secretariat for Research and  TMR No ERB FMAX-CT96-0090
of the European Union.


\begin{thebibliography}{99}

\bibitem{KTP} Kolb, E.W. and Turner M.S. (1990) 
{\it The Early Universe}, Addison, Wesley.\\
Peebles, P.J.E (1993), {\it Principles of Physical Cosmology}, 
Princeton University Press.\\
\bibitem{Jungm}For a recent review see e.g.
Jungman, G. {\it et al.}, (1996) {\it Phys. Rep.}{\bf 267}, 195.
\bibitem{COBE} Smoot, G.F. {\it et al.} (1992), 
{\it Astrophys. J.}{\bf 396} L1.
\bibitem{GAW}Gawiser E. and Silk J. (1988)
{\it {Science}} {\bf 280}, 1405;
Gross, M.A.K, Somerville, R.S., Primack, J.R., Holtzman, J. 
and Klypin, A.A. (1998)
{\it Mon. Not. R. Astron. Soc.} {\bf 301}, 81.
\bibitem{HSST}Riess, A.G. {\it et al} (1998), {\it Astron. J.}
{\bf 116}, 1009.
\bibitem{SPF} Somerville, R.S., Primack, J.R. and  Faber, S.M.(2000)
{\it astro-ph}/9806228;
{\it Mon. Not. R. Astron. Soc. (in press)}.
\bibitem{SCP}Perlmutter, S. {\it et al} (1999) {\it Astrophys. J.}
{\bf 517},565; (1997) {\bf 483},565.\\
Perlmutter, S., Turner, M.S. and White, M. (1999), 
{\it Phys. Rev. Let.} {\bf 83},670 (1999).
\bibitem{Turner} M.S. Turner, M.S. (1999)
{\it Cosmological parameters},{\it astro-ph}/9904051; (1990)
{\it Phys. Rep.} {\bf 333-334}, 619.
\bibitem{Benne} Bennett, D.P., {\it et al.} (1995),
 (MACHO collaboration), 
{\it A binary
lensing event toward the LMC: Observations and Dark Matter Implications, 
Proc. 5th Annual Maryland Conference, edited by S. Holt} ;
Alcock, C., {\it et al.}, (MACHO collaboration), 
{\it Phys. Rev. Lett.},{\bf 74} 2967. 
\bibitem{BERNA2} Bernabei, R. {\it et al}, (1998) INFN/AE-98/34;
Bernabei, R.,{\it et al} (19996), {\it Phys. Lett.} {\bf B 389}, 757.
\bibitem{BERNA1} Bernabei, R., { et al} (1998), {\it Phys. Lett.}
 {\bf B 424}, 195 ;
(1999) {\it Phys. Lett.}{\bf B 450}, 448.
\bibitem{JDV} Vergados,J.D. (1996) {\it J. of Phys.} {\bf G 22}, 253).
\bibitem{KVprd}Kosmas, T.S. and Vergados, J.D (1997), 
{\it Phys. Rev.} {\bf D 55}, 1752.
\bibitem{Dree}Drees, M. and Nojiri, M.M (1993)
{\it Phys. Rev.} {\bf D 48}, 3843;
\bibitem{Dree00}Djouadi, A. and Drees, M. (2000),
 {\it QCD Corrections to the Neuttalino
Nucleon Scattering}, TUM-HEP-370-00;PM/00-14;
Dawson, S. (1991), {\it Nucl. Phys.} {\bf B359},283;
Spira, M., {\it et al} (1995), {\it Nucl. Phys.} {\bf B453},17.
\bibitem{ref1}Goodman, M.W. and Witten, E. (1985), 
{\it Phys. Rev.} {\bf D 31}, 3059 (1985);
Griest, K. (1998), {\it Phys. Rev. Lett} {\bf  61}, 666; 
(1988), {\it Phys. Rev.} {\bf D 38}, 2357  ; 
(1989), {\it Phys. Rev.}{\bf D 39}, 3802;
Ellis, J. and Flores, R.A. (1991) 
{\it Phys. Lett.} {\bf B 263}, 259;
(1993), {\it Phys. Lett.} {\bf B 300}, 175; 
(1993),{\it Nucl. Phys.}{\bf B 400}, 25;
Ellis, J. and Roszkowski, L. (1992), 
{\it Phys. Lett.} {\bf B 283}, 252.
\bibitem{JDV99} Vergados, J.D. (1999) 
{\it Phys. Rev. Let.} {\bf 83}, 3597. 
\bibitem{JDV99b} Vergados J.D. (2000), 
{\it Phys. Rev.} {\bf D 62},023519-1; 
{\it astro-ph}/0001190
\bibitem{Gomez} M. Gomez and J.D. Vergados, 
(2000), {\it Direct LSP-Nucleus cross section} (to appear). 
\bibitem{ref2}Bottino, A., {\it et al.} (1992), 
{\it Mod. Phys. Lett.} {\bf A 7}, 733 (1992);
(1991), {\it Phys. Lett.} {\bf B 265}, 57; 
(1997), {Phys. Lett} {\bf B 402}, 113;
{\it hep-ph} $/$ 9709222; {\it hep-ph} $/$9710296;
Edsjo, J. and Gondolo, P. (1997), {\it Phys. Rev.} {\bf D 56}, 1789;
Berezinsky, Z., {\it et al.} (1996), 
{\it Astroparticle Phys.} {\bf 5}, 1;
Bednyakov, V.A., Klapdor-Kleingrothaus, H.V. and Kovalenko S.G.,
(1994), {\it Phys. Lett.} {\bf B 329}, 5.
\bibitem{ref3} Kane, G.L., {\it et al.} (1994),
{\it Phys. Rev.} {\bf D 49}, 6173 ;
Casta\v no, D.J., Piard , E.J. and Ramond, P. (1994), {\it Phys. Rev.} 
{\bf D 49}, 4882;  D.J. Casta\v no, {\it Private Communication};\\ 
Chamseddine, A.H., Arnowitt, R. and Nath, P. (1982), 
{Phys. Rev. Lett.} {\bf 49}, 970;
Nath, P, Arnowitt, R and Chamseddine, A.H. (1983),
 {\it Nucl. Phys.} {\bf B 227}, 121 ;
Arnowitt, R. and Nath, P. (1995),
{Mod. Phys. Lett.} {\bf 10}, 1215;
Arnowitt, R. and Nath, P. (1995), 
{\it Phys. Rev. Lett.}  {\bf 74}, 4952;
Arnowitt, R. and Nath, P. (1996), {\it Phys. Rev.} {\bf D 54}, 2394.
\bibitem{ref4} Arnowitt, R. and Nath, P. (1999),
{\it hep-ph}$/9701301$; {\it hep-ph}/9902237;
 Soni, S.K. and  Weldon, H.A. (1983), 
{\it Phys. Lett.} {\bf B 126}, 215
Kapunovsky, V.S. and Louis, J. (1993)
 {\it Phys. Lett.} {\bf B 306}, 268;
 (1986), {\it Phys. Lett} {\bf B 181}, 279 ;
Nath, P. and Arnowitt, R. (1989), {\it  Phys. Rev.} {\bf D 39}, 279;
Hagelin, J.S. and Kelly, S. (1990),
 {\it Nucl. Phys.} {\bf B 342}, 95;
Kamamura, Y, Murayama, H. and Yamaguchi, M (1994), 
{\it Phys. Lett.} {\bf B 324}, 52;
Dimopoulos S. and Georgi, H. (1981), {\it Nucl. Phys.} {\bf B 206}, 387.
\bibitem{Hab-Ka} Haber, H.E. and Kane, G.L. (1985),
Phys. Rep. {\bf  117} 75 (1985).
\bibitem{Chen} Cheng, T.P. (1998), {\it Phys. Rev.} {\bf D 38} 2869;
Cheng, H.-Y. (1989), {\it Phys. Lett.} {\bf B 219} 347.
\bibitem{Ress}Ressell, M.T., {\it et al.} (1993),
 {\it Phys. Rev}. {\bf D 48}, 5519;
Ressell, M.T. and Dean, D.J. (1997), {\it hep-ph}$/9702290$.
\bibitem{Dimit}Dimitrov, V.I., Engel, J. and Pittel, S. (1995),
{\it Phys. Rev.} {\bf D 51}, R291.
\bibitem{Engel}Engel, J. (1991), {\it Phys. Lett.} {\bf B 264}, 114.
\bibitem{Nikol}Nikolaev, M.A. and  Klapdor-Kleingrothaus, H.V. (1993),
{\it Z. Phys.} 
{\bf A 345}, 373; (1993), {\it Phys. Lett.} {\bf 329 B}, 5; 
(1995), {\it Phys. Rev.} {\bf D 50}, 7128.
\bibitem{DIVA00} Divari, P.C., Kosmas, T.S., Vergados, J.D. and 
Skouras, L.D. (2000),
{\it Phys. Rev.} {\bf C 61}, 044612-1.
\bibitem{Copi99}Copi, C.J., Heo, J. and Krauss, L.M. (1999),
{\it Phys. Lett.} {\bf 461 B}, 43; 
\bibitem{Druk}Drukier, A.K.,{\it et al}, (1986) 
{\it Phys. Rev.} {\bf D 33}, 3495;\\
Collar, J.I., {\it et al} (1992), {\it Phys. Lett} {\bf B 275}, 181.
\bibitem{SIKIVIE}Sikivie, P., Tkachev, I. and Wang, Y., (1995)
{\it  Phys. Rev. Let.} {\bf 75}, 
2911 ; (1997), {\it Phys. Rev.} {\bf D 56}, 1863.
Sikivie, P. (1998), {\it Phys. Let.} {\bf B 432}, 139; astro-ph/9810286 
\bibitem{Smith}Smith, P.F., {\it et al} (1999), 
{\it Phys. Rep.} {\bf 307}, 275 (1999); 
Spooner, N. (1999), {\it Phys.Rep.} {\bf 307}, 253;
Quenby, J.J., {\it et al} (1996) {\it Astropart. Phys.} {\bf 5}, 249).
\bibitem{Primack}Primack, J.R., Seckel, D. and Sadoulet, B. (1998)
{\it Ann. Rev. Nucl. Part. Sci.} {\bf 38}, 751.
\bibitem{Smith1} Smith, P.F. and Lewin, J.D. (1990) 
{\it Phys. Rep.} {\bf 187}, 203. 
\bibitem{KVdubna}Vergados, J.D. and Kosmas, T.S. (1998), 
{\it Physics of Atomic nuclei},
 Vol. {\bf 61}, No {\bf 7}, 1066  (from {\it Yadernaya Fisika}, 
Vol.{\bf 61}, No{\bf 7}, 1166.
\bibitem{Verg98}Vergados, J.D. (1998),
 {\it Phys. Rev.} {\bf D 58}, 103001-1.
\end{thebibliography}
\end{document}